\documentclass[aps,prd,onecolumn,showkeys,showpacs,amssymb]{revtex4}
\usepackage{float}
\usepackage{amssymb}
\usepackage{graphicx}
\usepackage{amsmath}
\usepackage{hyperref}
\begin{document}

\title{Magnetic Neutron Stars in $f(R)$  gravity}

\author{Artyom V. Astashenok$^{1}$, Salvatore Capozziello$^{2,3,4}$, Sergei D. Odintsov$^{5,6,7}$}

\affiliation{$^{1}$I. Kant Baltic Federal University, Institute of Physics and Technology, Nevskogo st. 14, 236041 Kaliningrad, Russia.\\
$^2$Dipartimento di Fisica, Universita' di Napoli "Federico II" and
\\$^3$INFN Sez. di Napoli, Complesso Universitario di Monte S. Angelo, Ed.
G., Via Cinthia,
9, I-80126, Napoli, Italy,\\
$^4$ Gran Sasso Science Institute (INFN), Viale F. Crispi, 7, I-67100. L'Aquila, Italy.\\
$^5$Instituci\`{o} Catalana de Recerca i Estudis Avan\c{c}ats (ICREA), Barcelona, Spain,\\
$^6$Institut de Ciencies de l'Espai (IEEC-CSIC), Campus UAB, Torre C5-Par-2a pl, E-08193 Bellaterra, Barcelona, Spain,\\
$^7$Tomsk State Pedagogical University (TSPU), Tomsk, Russia.}

\begin{abstract}
Neutron stars with strong magnetic fields are considered in
the framework of $f(R)$ gravity. In order to describe dense
matter in magnetic field, the model with baryon octet
interacting through $\sigma$$\rho$$\omega$-fields is used. The
hyperonization process results in softening  the equation of state (EoS)
and in   decreasing the maximal mass. We investigate the effect of
strong magnetic field in  models involving quadratic and cubic corrections in the Ricci scalar $R$ to the Hilbert-Einstein action. For large fields,
the Mass-Radius relation differs considerably from that of  General
Relativity only for stars with masses close to the maximal one. Another
interesting feature is the  possible existence of more compact stable
stars with extremely large magnetic fields ($\sim 6\times 10^{18}$ G
instead of $\sim 4\times 10^{18}$ G  as in General Relativity) in the central
regions of the stars. Due to cubic terms, a  significant increasing of the
maximal mass is possible.

\end{abstract}

\keywords{magnetars, neutron stars, modified gravity.}

\pacs{11.30.-j; 97.60.Jd; 04.50.Kd}

\date{\today}

\maketitle

\section{Introduction}

Neutron stars are observed as  several classes of self-gravitating systems: as radio and X-ray pulsars, as X-ray
bursters, as compact thermal X-ray sources in supernova remnants,
as rotating radio transients. In general, the structure of neutron stars and  the
relation between the mass and the radius are determined by equations of
state (EoS) of dense matter.

The maximal mass of neutron star is still an open question. Recent
observations allows to estimate this limit at least as
$2M_{\odot}$: the well-measured limit  of the pulsar PSR J1614-2230 is
$1.97M_{\odot}$ \cite{Demorest}, while, for the  pulsar J0348+0432, it  is
$2.01M_{\odot}$ \cite{Antoniadis}. Other examples of massive
neutron stars are Vela X-1 ($\sim 1.8M_{\odot}$ \cite{Rawls}) and
4U 1822-371 ($\sim 2M_{\odot}$, \cite{Munoz}).  There are some
indications in favor of  the existence of more massive neutron stars
with masses $\sim 2.4M_{\odot}$ (the possible masses of B1957+20
\cite{Kerk} and 4U 1700-377 \cite{Clark}) or even $\sim
2.7M_{\odot}$ (J1748-2021B \cite{Freire}).

It is interesting to note that for various EoS including hyperons,
the maximal mass limit for non-magnetic neutron stars is
considerably below than of the two-solar masses limit. The hyperonization process softens
EoS and then the  maximal  allowable mass results  reduced
\cite{Glendenning,Glendenning-2,Schaffner,Vidana,Schulze}.

There are several ways to approach  the solution of this problem (the so called ``{\it hyperon  puzzle}'').

Firstly, the extensions of the simple model of hyperonic matter
(with three exchange meson fields - the so called
'$\rho\omega\sigma$'-model) allow to achieve the increasing of
maximal mass. Various approaches  are proposed following this track. For
instance,   larger hyperon-vector couplings (in comparison with
quark counting rule) require  stiffness of the EoS
\cite{Hofmann,Rikovska,Sedrakian,Miyatsu-2}. Similar effect occurs
in model with chiral quark-meson coupling \cite{Miyatsu}. The
quartic vector-meson terms in the Lagrangian \cite{Bednarek} or the
inclusion of an additional vector-meson mediating repulsive
interaction amongst hyperons \cite{Weissenborn} also lead to the
increasing of the maximal mass limit. Authors of Ref. \cite{Whittenbury}
proposed an EoS with maximum mass $\sim 2.1M_{\odot}$ using the
quark-meson coupling model, which naturally incorporates hyperons
without additional parameters. A model with in-medium hyperon
interactions is considered in \cite{Wei}.

Another source for  increasing the maximal mass  limit  is the
existence of strong magnetic fields inside the star.
The existence of soft gamma-ray repeaters and anomalous X-ray
pulsars can be linked to  neutron stars with very
strong magnetic fields of the order $10^{15}$ G on the surface.
In these cases, the maximum magnetic field in the central  regions of neutron star can
exceed $10^{18}$ G, according to the scalar virial
theorem. Such magnetic fields affect considerably the  EoS for dense
matter and result in  increasing  the maximal mass of neutron
stars.

Various models of dense nuclear matter in  presence of
strong magnetic fields have been considered in literature. The simplest model with
interacting $npe\mu$ gas is investigated in \cite{Lattimer}.
Models with hyperons and quarks are considered in
\cite{Rabhi}-\cite{Lopes}. It has been demonstrated  that the Landau
quantization leads to the softening of the EoS for matter but account
for  contributions of magnetic field into pressure and density.  This fact  leads, on the other hand,
to the  stiffening of  EoS.

Therefore neutron stars are very peculiar objects for testing theories
of matter at high density regimes  and in strong magnetic fields.
It is interesting to note that data about neutron stars (mainly
mass-radius ($M-R$) relation) can be used for investigating
possible deviations  from General Relativity (GR).

The initial motivation for studying  modified gravity came from the
 discovered accelerated expansion of the universe
confirmed by numerous independent observations. These observations
include Hubble diagram for Ia type supernovae \cite{Perlmutter,
Riess1,Riess2}, cosmic microvawe background radiation (CMBR) data
\cite{Spergel}, surveys of gravitational weak lensing
\cite{Schmidt} and data on Lyman alpha forest absorption lines
\cite{McDonald}.

This acceleration takes place at relatively small distances
(``Hubble flow'') and requires (in GR)
non-standard cosmic fluid (dark energy) filling the  universe  with negative pressure but  not
clustered in large scale structure. The
nature of dark energy is unclear.  Although from an observational
viewpoint, the so called $\Lambda$CDM model (where  dark
energy is considered as Einstein Cosmological Constant) is in
agreement with data coming from observations there are various
problems and shortcomings at  theoretical level.  One of this issues is the
``smallness'' of cosmological constant i.e. the difference of 120
orders of magnitude between its observed value and the one
predicted by quantum field theory \cite{Weinberg}.

An alternative approach to dark energy problem consists of
extending of GR. In this case,  the accelerated
expansion can be obtained without using ``dark energy'' but
enlarging the gravitational sector \cite{Capozziello1,
Capozziello2, Odintsov1, Turner, Odintsov-3,
Capozziello3,Capozziello_book, Capozziello4, Cruz}. Therefore
theories of modified gravity can be considered as real alternative
to GR.

The study of relativistic stars in modified gravity is interesting
from several reasons and could constitute a formidable probe for
such theories. Firstly one can reject some models that  do not
allow the existence of stable star configurations \cite{Briscese,
Abdalla, Bamba, Kobayashi-Maeda, Nojiri5,Lang} (however one has to
note that stability can be achieved due to ``chameleon mechanism''
\cite{Tsujikawa, Upadhye-Hu} and may depend on the choice of the
EoS). Secondly there is the possibility  for the existence of
 new stellar structures, in the framework of modified gravity, escaping the standard stellar models.
The observation of such self-gravitating  anomalous structures could provide strong
evidence for the Extended Gravity (see e.g. \cite{Laurentis, Laurentis2, Farinelli}).

The present  paper is devoted to neutron stars with strong magnetic fields
in framework  of analytic $f(R)$ gravity. Assuming a
simple model for strong interactions, one can obtain the EoS for
dense matter in magnetic field. Landau quantization, due to
magnetic fields, results to have significant effects. We consider the
cases of slowly and fast varying fields.

The paper is organized as follows. In Sec. II, we briefly
consider the the field equations for $f(R)$ gravity and the
modified Tolman--Oppenheimer--Volkoff (TOV) equations. Then
relativistic mean field theory for dense matter in strong magnetic
fields is presented (Sec.III).

In Sec. IV, the neutron star models for strong magnetic fields in
 quadratic ($f(R)=R+\alpha R^2$) and cubic $f(R)=R+\alpha
R^3$ gravity are presented. The $M-R$ relation is derived and
compared with the one in GR. Conclusions and
outlooks are reported  in  Sec.V.

\section{Modified TOV equations in $f(R)$ gravity}
The action of $f(R)$-gravity is
\begin{equation}\label{action}
S=\frac{c^4}{16\pi G}\int d^4x \sqrt{-g}f(R) + S_{{\rm
matter}}\quad\,.
\end{equation}
It can be expressed as $f(R)=R+\alpha h(R)$.
The field equations are
\begin{equation}\label{field}
(1+\alpha h_{R})G_{\mu \nu }-\frac{1}{2}\alpha(h-h_{R}R)g_{\mu \nu
}-\alpha (\nabla _{\mu }\nabla _{\nu }-g_{\mu \nu }\Box
)h_{R}=8\pi G T_{\mu \nu }/c^{4}.
\end{equation}
Here $g$ is the determinant of the metric $g_{\mu\nu}$ and $S_{\rm
matter}$ is the action of the standard perfect fluid matter. The
Einstein tensor is $G_{\mu\nu}=R_{\mu\nu}-\frac12Rg_{\mu\nu}$ and
${\displaystyle h_R=\frac{dh}{dR}}$.

For  star configurations,  one can assumes a spherically symmetric
metric with two independent functions of radial coordinate, that
is:
\begin{equation}\label{metric}
    ds^2= -e^{2\phi}c^2 dt^2 +e^{2\lambda}dr^2 +r^2 (d\theta^2
    +\sin^2\theta d\phi^2).
\end{equation}
For the exterior solution, we assume a Schwarzschild metric.
Therefore it is convenient to define the variable \cite{Stephani,Cooney}
\begin{equation}\label{mass}
    e^{-2\lambda}=1-\frac{2G M}{c^2 r}.
\end{equation}
The value of variable $M$ on the star surface is the gravitational
mass. For a perfect fluid, the energy-momentum tensor is
$T_{\mu\nu}=\mbox{diag}(e^{2\phi}\rho c^{2}, e^{2\lambda}P, r^2P,
r^{2}\sin^{2}\theta P)$,  where $\rho$ is the matter density and $P$ is
the pressure. The field equations of interest  are
\begin{eqnarray}
  -8\pi G \rho/c^2 &=& -r^{-2} +e^{-2\lambda}(1-2r\lambda')r^{-2}
                +\alpha h_R(-r^{-2} +e^{-2\lambda}(1-2r\lambda')r^{-2}) \nonumber \\
             && -\frac12\alpha(h-h_{R}R) +e^{-2\lambda}\alpha[h_R'r^{-1}(2-r\lambda')+h_R''] \label{f-tt},\\
  8\pi G P/c^4 &=& -r^{-2} +e^{-2\lambda}(1+2r\phi')r^{-2}
                +\alpha h_R(-r^{-2} +e^{-2\lambda}(1+2r\phi')r^{-2}) \nonumber \\
             && -\frac12\alpha(h-h_{R}R) +e^{-2\lambda}\alpha h_R'r^{-1}(2+r\phi'), \label{f-rr}
\end{eqnarray}
where $'\equiv d/dr$. The second TOV equation follows from the conservation law
$T_{\nu;\mu}^{\mu}=0$ and Eq.(\ref{f-rr}).
As result, the modified TOV equations can be written as
\cite{Astashenok}
\begin{equation}\label{TOV-1}
\left(1+\alpha  h_{{R}}+\frac{1}{2}\alpha h'_{{R}}
r\right)\frac{dm}{dr}=4\pi{\rho}r^{2}-\frac{1}{4}\alpha r^2
\left[h-h_{{R}}{R}-2\left(1-\frac{2m}{r}\right)\left(\frac{2h'_{{R}}}{r}+h''_{{R}}\right)\right],
\end{equation}
\begin{equation}\label{TOV-2}
8\pi p=-2\left(1+\alpha
h_{{R}}\right)\frac{m}{r^{3}}-\left(1-\frac{2m}{r}\right)\left(\frac{2}{r}(1+\alpha
h_{{R}})+\alpha r_{g}^{2}
h'_{{R}}\right)({\rho}+p)^{-1}\frac{dp}{dr}-
\end{equation}
$$
-\frac{1}{2}\alpha
\left[h-h_{{R}}{R}-4\left(1-\frac{2m}{r}\right)\frac{h'_{{R}}}{r}\right],
$$
Here we have  introduced the dimensionless variables $M=m M_{\odot},\quad
r\rightarrow r_{g}r, \quad \rho\rightarrow\rho
M_{\odot}/r_{g}^{3},\quad P\rightarrow p M_{\odot}c^{2}/r_{g}^{3},
\quad R\rightarrow {R}/r_{g}^{2}$, $\alpha
r_{g}^{2}h(R)\rightarrow \alpha h(R)$, where
$r_{g}=GM_{\odot}/c^{2}=1.47473$ km.
The third independent equation for Ricci curvature scalar is
\begin{equation}\label{TOV-3}
3\alpha
r_{g}^{2}\left[\left(\frac{2}{r}-\frac{3m}{r^{2}}-\frac{dm}{rdr}-\left(1-\frac{2m}{r}\right)\frac{dp}{(\rho+p)dr}\right)\frac{d}{dr}+
\left(1-\frac{2m}{r}\right)\frac{d^{2}}{dr^{2}}\right]h_{{R}}+\alpha
r_{g}^{2} h_{{R}}{R}-2\alpha r_{g}^{2} h-{R}=-8\pi({\rho}-3p)\,.
\end{equation}
 Eqs. (\ref{TOV-1}), (\ref{TOV-2}), and  (\ref{TOV-3}) can be solved
numerically for given EOS. In order to get  solution,  one can use perturbative
approach (see for details
\cite{Arapoglu,Alavirad,Astashenok,Astashenok-2}). In the framework  of
perturbative approach,  terms containing $h(R)$ are assumed to
be of first order in the small parameter $\alpha$, so all such
terms should be evaluated at ${\mathcal O}(\alpha)$ order. The
Ricci curvature scalar at zero order is $R^{(0)}=8\pi
(\rho^{(0)}-3p^{(0)})$. Therefore the deviation from GR strongly depends from the assumed form of EoS.

\section{Relativistic mean field theory for dense matter in presence of strong magnetic field}

Let us assume a simple model for describing nuclear matter in magnetic
field. The magnetic field $B$ is assumed along $z$-axis i.e. the
4-potential is $A^{mu}=(0,0,Bx,0)$. For nuclear matter consisting
of baryon octet ($b=$$p$, $n$, $\Lambda$, $\Sigma^{0,\pm}$,
$\Xi^{0,-}$) interacting with magnetic field and scalar $\sigma$,
isoscalar-vector $\omega_\mu$ and isovector-vector $\rho_\mu$ meson
fields and leptons ($l=$$e^{-}$, $\mu^{-}$), it  is \cite{Typel}
\begin{equation}
\mathcal{L}=\sum_{b}\bar{\psi}_{b}\left(\gamma_{\mu}(i\partial^{\mu}-q_{b}A^{\mu}-g_{\omega
b}\omega^{\mu}-\frac{1}{2}g_{\rho
b}{\tau}\cdot{\rho}^{\mu})-(m_{b}-g_{\sigma
b}\sigma)\right)\psi_{b}+\sum_{l}\bar{\psi}_{l}\left(\gamma_{\mu}(i\partial^{\mu}-q_{l}A^{\mu})-m_{l}\right)\psi_{l}+
\end{equation}
$$
+\frac{1}{2}\left((\partial_{\mu}\sigma)^{2}-m^{2}_{\sigma}\sigma^{2}\right)-V(\sigma)-\frac{1}{4}F_{\mu\nu}F^{\mu\nu}+\frac{1}{2}m^{2}_{\omega}\omega^{2}-\frac{1}{4}\omega_{\mu\nu}\omega^{\mu\nu}-
\frac{1}{4}{\rho}_{\mu\nu}{\rho}^{\mu\nu}+\frac{1}{2}m^{2}_{\rho}{\rho}_{\mu}^{2}.
$$
Here the mesonic and electromagnetic field strength tensors are
defined by the usual relations
$\omega_{\mu\nu}=\partial_{\mu}\omega_{\nu}-\partial_{\nu}\omega_{\mu}$,
$\rho_{\mu\nu}=\partial_{\mu}\rho_{\nu}-\partial_{\nu}\rho_{\mu}$,
$F_{\mu\nu}=\partial_{\mu}A_{\nu}-\partial_{\nu}A_{\mu}$. For
the sake of simplicity,  we consider frozen-field configurations of
electromagnetic field. Also we neglect the anomalous magnetic
moments (AMM) of baryons and leptons because their effect is very
small. The strong interaction couplings $g_{b\sigma}$,
$g_{b\omega}$ and $g_{b\rho}$ depend from density. We use the
parameterization adopted in \cite{Typel}:

\begin{equation}
g_{i}(\rho)=g_{i0}f_{i}(x), x=\rho/\rho_{0}\,,
\end{equation}
where

$$f_{i}(x)=a_{i}\frac{1+b_{i}(x+d_{i})^2}{1+c_{i}(x+d_{i})^{2}}\,.$$
For the  isovector field, it is
$$
g_{b\rho}=g_{b0}\exp[-a_{\rho}(x-1)].
$$
The values of constants $a_{i}$, $b_{i}$, $c_{i}$, $d_{i}$ are
given in \cite{Typel}. Using the mean-field approximation,  one can obtain the following
equations for meson fields:
\begin{equation}\label{0}
m^{2}_{\sigma}\sigma+\frac{dV}{d\sigma}=\sum_{b}g_{\sigma b}
n_{b}^{s},\quad m^{2}_{\omega}\omega_{0}=\sum_{b}g_{\omega b}
n_{b},\quad m^{2}_{\omega}\rho_{03}=\sum_{b}g_{\rho b} n_{b}.
\end{equation}
Here $\sigma$, $\omega^{0}$, $\rho^{0}$ are the expectation values
of the meson fields in uniform matter. The quantities $n^{s}_{b}$,
$n_{b}$ are the scalar and vector baryon number densities,
correspondingly. The simplest scalar field potential is defined as
\begin{equation}
V(\sigma)=\frac{1}{3}b m_{N}(g_{\sigma
N}\sigma)^{3}+\frac{1}{4}c(g_{\sigma N}\sigma)^{4},
\end{equation}
where $b$ and $c$ are dimensionless constants. The values of
nucleon-meson couplings and parameters $b$, $c$ are given in Table
I. From the Dirac equations for charged and neutral baryons and
leptons,  we have the energy spectra:
\begin{equation}
E^{b}_{\nu}=(k_{z}^{2}+m_{b}^{*2}+2\nu|q_{b}|B)^{1/2}+g_{\omega
b}\omega^{0}+\tau_{3b}g_{\rho b}\rho^{0}+\Sigma^{R}_{0},
\end{equation}
\begin{equation}
E^{b}=(k^{2}+m_{b}^{2})^{1/2}+g_{\omega
b}\omega^{0}+\tau_{3b}g_{\rho b}\rho^{0}+\Sigma^{R}_{0},
\end{equation}
\begin{equation}
E^{l}_{\nu}=(k_{z}^{2}+m_{l}^{2}+2\nu|q_{l}|B)^{1/2}.
\end{equation}
The number $\nu=n+1/2-sgn(q)s/2$ denotes the Landau levels of the
fermions with electric charge $q$, spin number $s=\pm 1$ for spin
up and spin down cases correspondingly. The spin degeneracy is
$g_{\nu}=1$ for lowest Landau level ($\nu=0$) and 2 for all other
levels. The effective mass for baryons is $m_{b}^{*}=m_{b}-g_{\sigma
b}\sigma$.

The rearrangement self-energy term is defined by
\begin{equation}
\Sigma^{R}_{0}=-\frac{\partial \ln g_{\sigma N}}{\partial
n}m^{2}_{\sigma}\sigma^{2}+\frac{\partial \ln g_{\omega
N}}{\partial n}m^{2}_{\omega}\omega^{2}_{0}+\frac{\partial \ln
g_{\rho N}}{\partial n}m^{2}_{\rho}\rho^{2}_{0}.
\end{equation}
Here $n=\sum_{b} n_{b}$.

The scalar densities for neutral baryons are \cite{Lattimer}
\begin{equation}
n^{s}_{b}=\frac{m_{b}^{*2}}{2\pi^2}\left(E^{b}_{f}k^{b}_{f}-
m_{b}^{*2}\ln\left|\frac{k^{b}_{f}+E^{b}_{f}}{m_{b}^{*}}\right|\right)
\end{equation}
and for charged baryons, it is
\begin{equation}
n^{s}_{b}=\frac{|q_{b}|B m_{b}^{*}}{2\pi^2}\sum_{\nu}
g_{\nu}\ln\left|\frac{k^{b}_{f,\nu}+E^{b}_{f}}{\sqrt{m_{b}^{*2}}+2\nu|q_{b}|B}\right|.
\end{equation}
For the vector densities for neutral baryons, we have
\begin{equation}
n_{b}=\frac{1}{3\pi^2} k^{b 3}_{f}.
\end{equation}
and for charged baryons and leptons, it is
\begin{equation}
n_{b,l}=\frac{|q_{b,l}|B}{2\pi^2} \sum_{\nu} g_{\nu}
k^{b,l}_{f,\nu}.
\end{equation}

Here $E_{f}^{b,l}$ is the Fermi energy.  For charged baryon,
$E_{f}^{b}$  is related to the Fermi momentum $k_{f,\nu}^{b}$ as
$E_{f}^{b}=(k_{f}^{2}+m^{*2}_{b}+2\nu |q_{b}| B)^{1/2}$.  For
neutral baryon, it is  $E_{f}^{b}=(k_{f}^{2}+m^{*2}_{b})^{1/2}$. The
summation over $\nu$ terminates at value $\nu_{max}$ where the
square of Fermi momenta is still positive. For large magnetic
fields $B\sim 10^{18}$ G,  only few Landau levels are occupied.

For hyperon-meson couplings there are no well-defined rule. One
can use for these constants quark counting rule \cite{Dover,Schafner}:
\begin{equation}
g_{\omega \Lambda}=g_{\omega \Sigma}=\frac{2}{3}g_{\omega N},
\quad g_{\omega \Xi}=\frac{1}{2}g_{\omega N},
\end{equation}
and
\begin{equation}
g_{\rho \Sigma}=2g_{\rho N}, \quad g_{\rho\Xi}=g_{\rho N}.
\end{equation}

Another choice is assuming that the fractions of
nucleon-meson couplings, i.e. $g_{iH}=x_{iH}g_{iN}$, is fixed.  Here, it is
$x_{\sigma H}=x_{\rho H}=0.600$, $x_{\omega H}=0.653$, $x_{\rho
H}=0.6$ (see \cite{Rabhi}). We use this definition for further
calculations.

\begin{table}
\label{Table1}
\begin{centering}
\begin{tabular}{|c|c|c|c|c|c|c|c|c|}
  \hline
   & $n_{s}$,  & $-B/A$, & $M^{*}/M$ & $g_{0\sigma N}/m_{\sigma}$, & $g_{0\omega N}/m_{\omega}$, & $g_{0\rho N}/m_{\rho}$, &  &  \\
  Model & fm$^{-3}$ & MeV & & fm & fm & fm & b & c \\
  \hline
  TW & 0.153 & 16.30 & 0.56 & 3.84901 & 3.34919 & 1.89354 & 0 & 0 \\
  GM1 & 0.153 & 16.30 & 0.70 & 3.434 & 2.674 &  2.100 & 0.002947 & $-0.001070$ \\
  GM2 & 0.153 & 16.30 & 0.78 & 3.025 & 2.195 &  2.189 & 0.003487 & $0.01328$ \\
  GM3 & 0.153 & 16.30 & 0.78 & 3.151 & 2.195 &  2.189 & 0.008659 & $-0.002421$ \\
  \hline
\end{tabular}
\caption{The nucleon-meson couplings and parameters of scalar
field potential for some models (GM1-3 -  \cite{Glendenning}, TW -
\cite{Typel}). The nuclear saturation density $n_{s}$, the Dirac
effective mass $M^{*}$ and the binding energy ($B/A$) are also
given.}
\end{centering}
\end{table}

For chemical potential of baryons and leptons,  one has
$$
\mu_{b}=E^{f}_{b}+g_{\omega b} \omega_{0}+g_{\rho
b}I_{3}\rho_{03}+\Sigma_{0}^{R},\quad \mu_{l}=E^{f}_{l}.
$$
 The following conditions should be
imposed on the matter in order to  obtain the EoS with the following properties:\\
\\ (i) baryon number
conservation:
\begin{equation}\label{1}
\sum_{b}n_{b}=n,
\end{equation}
(ii) charge neutrality:
\begin{equation}\label{2}
\sum_{i} q_{i}n_{i}=0,\quad i=b,l,
\end{equation}
(iii) beta-equilibrium conditions:
\begin{equation}\label{3}
\mu_{n}=\mu_{\Lambda}=\mu_{\Xi^{0}}=\mu_{\Sigma^{0}}, \quad
\mu_{p}=\mu_{\Sigma^{+}}=\mu_{n}-\mu_{e},\quad
\mu_{\Sigma^{-}}=\mu_{\Xi^{-}}=\mu_{n}+\mu_{e},\quad
\mu_{e}=\mu_{\mu}.
\end{equation}
At given $n$,  Eqs. (\ref{0})-(\ref{3}) can be solved
numerically and one can find the Fermi energy for particles
and meson fields. Resulting matter energy density is
\begin{equation}
\epsilon_{m}=\sum_{b} \epsilon_{b}+\sum_{l}
\epsilon_{l}+\frac{1}{2}m_{\sigma}^{2}+\frac{1}{2}m_{\omega}^{2}+\frac{1}{2}m_{\rho}\rho^{2}_{0}+U(\sigma).
\end{equation}
The energy density for charged baryons is
\begin{equation}\label{encb}
\epsilon^{c}_{b}=\frac{|q_{b}|B}{4\pi^2}\sum_{\nu}g_{\nu}\left(k^{b}_{f,\nu}E^{b}_{f}+
(m_{b}^{*2}+2\nu|q_{b}|B)\ln\left|\frac{k^{b}_{f,\nu}+E^{b}_{f}}{\sqrt{m_{b}^{*2}+2\nu|q_{b}|B}}\right|\right)
\end{equation}
and for neutral baryons
$$
\epsilon^{n}_{b}=\frac{1}{4\pi^2}\left[k^{b}_{f}(E^{b}_{f})^{3}-\frac{1}{2}m_{b}^{*}\left(m_{b}^{*}k^{b}_{f}E^{b}_{f}
+m_{b}^{*3}\ln\left|\frac{k^{b}_{f}+E^{b}_{f}}{m_{b}^{*}}\right|\right)\right].
$$
The expression of energy density for leptons can be obtained from
(\ref{encb}) by changing $m_{b}^{*}\rightarrow m_{l}$. The pressure
of dense matter is defined as
$$
p=\sum_{b} p_{b}+\sum_{l}
p_{l}-\frac{1}{2}m_{\sigma}^{2}+\frac{1}{2}m_{\omega}^{2}+\frac{1}{2}m_{\rho}\rho^{2}_{0}-U(\sigma)+\Sigma^{R}_{0}.
$$
where pressure for charged baryons is
\begin{equation}
p^{c}_{n}=\frac{|q_{b}|B}{12\pi^2}\sum_{\nu}g_{\nu}\left(k^{b}_{f,\nu}E^{b}_{f}-
(m_{b}^{*2}+2\nu|q_{b}|B)\ln\left|\frac{k^{b}_{f,\nu}+E^{b}_{f}}{\sqrt{m_{b}^{*2}+2\nu|q_{b}|B}}\right|\right)
\end{equation}
and for neutral baryons
\begin{equation}
p^{n}_{b}=\frac{1}{12\pi^2}\left[k^{b}_{f}(E^{b}_{f})^{3}-\frac{3}{2}m_{b}^{*}\left(m_{b}^{*}k^{b}_{f}E^{b}_{f}
-m_{b}^{*3}\ln\left|\frac{k^{b}_{f}+E^{b}_{f}}{m_{b}^{*}}\right|\right)\right].
\end{equation}

In order to obtain the  EoS,  one needs to add the contribution of magnetic
field, that is
\begin{equation}
\epsilon=\epsilon_{m}+\frac{B^2}{8\pi},\quad
p=p_{m}+\frac{B^2}{8\pi}.
\end{equation}
We use a model where the magnetic field
depends from baryon density only. The parameterization proposed in \cite{Rabhi}, \cite{Ryu-2} has the form
\begin{equation}
B=B_{s}+B_{0}\left[1-\exp\left(-\beta(n/n_{s})^{\gamma}\right)\right],
\end{equation}
where $B_s$ is the magnetic field on the star surface ($10^{15}$ G).
For parameters $\gamma$ and $\beta$,  one takes the values
$\gamma=2$, $\beta=0.05$ (slowly varying field) and $\gamma=3$,
$\beta=0.02$ (fast varying field). The value $B_{0}$ is convenient
to give in units of critical field for electron $B_{c}=4.414\times
10^{13}$ G.

All these considerations can be applied to models where curvature corrections appear in the TOV equations. Specifically, we adopt quadratic and cubic corrections.

\begin{table}
\label{Table2}
\begin{centering}
\begin{tabular}{|c|c|c|c|c|c|}
  \hline
  $B_{0}$,  & $\alpha$, & $M_{max}$, & $R$,  & $E_{c}$, & $B_{c}$, \\
  $10^5$ G & $10^{9}$ cm$^{2}$ & $M_{\odot}$ & km & GeV/fm$^{3}$ & $10^{18}$ G \\
  \hline
   & 0 & 1.51 & 10.00 & 1.61 & 0 \\
  0 & $-5$ & 1.55 & 10.00 & 1.61 & 0 \\
   & $5$ & 1.46 & 10.05 & 1.49 & 0 \\
   \hline
   & 0 & 2.21 & 11.69 & 1.17 & 3.38 \\
  1 & $-5$ & 2.30  & 11.58 & 1.27 & 3.56 \\
   & 5 & 2.14 & 11.82  & 1.09 & 3.20  \\
   \hline
      & 0 & 2.80 & 13.99 & 0.79 & 3.50 \\
  2 & $-5$ & 2.91 & 13.44 & 0.97 & 3.93 \\
   & $5$ & 2.71 & 14.31 & 0.68 & 3.06 \\
  \hline
    & 0 & 3.21 & 15.67 & 0.63 & 3.29 \\
  3 & $-5$ & 3.33 & 15.24 & 0.68 & 3.75 \\
   & 5  & 3.11 & 16.03 & 0.54 & 2.87 \\
   \hline
\end{tabular}
\caption{Neutron star properties using TW model for quadratic
gravity (slowly varying field). The energy density ($E_{c}$) and
magnetic field ($B_{c}$) in the center for neutron star with
maximal mass are given.}
\end{centering}
\end{table}

\begin{table}
\label{Table3}
\begin{centering}
\begin{tabular}{|c|c|c|c|c|c|}
  \hline
  $B_{0}$,  & $\alpha$, & $M_{max}$, & $R$,  & $E_{c}$, & $B_{c}$, \\
  $10^5$ & $10^{9}$ cm$^{2}$ & $M_{\odot}$ & km & GeV/fm$^{3}$ & $10^{18}$ G \\
  \hline
   & 0 & 2.32 & 11.50 & 1.17 & 3.96 \\
  1 & $-5$ & 2.44  & 11.47 & 1.22 & 4.36 \\
   & 5 & 2.23 & 11.66 & 1.04 & 3.66 \\
   \hline
      & 0 & 2.73 & 12.44 & 0.93 & 4.20 \\
  2 & $-5$ & 2.89 & 12.81 & 0.97 & 4.34 \\
   & $5$ & 2.60 & 13.31 & 0.71 & 3.29 \\
  \hline
    & 0 & 2.98 & 13.78 & 0.76 & 3.87 \\
  3 & $-5$ & 3.12 & 13.81 & 0.83 & 4.12 \\
   & 5  & 2.84 & 14.06 & 0.68 & 3.50 \\
   \hline
\end{tabular}
\caption{Neutron star properties using TW model for quadratic
gravity (fast varying field).}
\end{centering}
\end{table}

\begin{table}
\label{Table4}
\begin{centering}
\begin{tabular}{|c|c|c|c|c|c|}
  \hline
  $B_{0}$,  & $\alpha$, & $M_{max}$, & $R$,  & $E_{c}$, & $B_{c}$, \\
  $10^5$ & $10^{9}$ cm$^{2}$ & $M_{\odot}$ & km & GeV/fm$^{3}$ & $10^{18}$ G \\
  \hline
  2 & $10$ & 2.30 & 11.82 & 1.49 & 5.97 \\
  \hline
  3 & $10$ & 2.52 & 12.86 & 1.40 & 6.08 \\
   \hline
\end{tabular}
\caption{Compact star properties on the second ``branch of
stability'' using TW model for quadratic gravity (fast varying
field). The magnetic field at the center can exceed $6\times
10^{18}$ G and the central energy density is approximately twice
than in GR.}
\end{centering}
\end{table}

\begin{center}
\begin{figure}
  \includegraphics[scale=1.1]{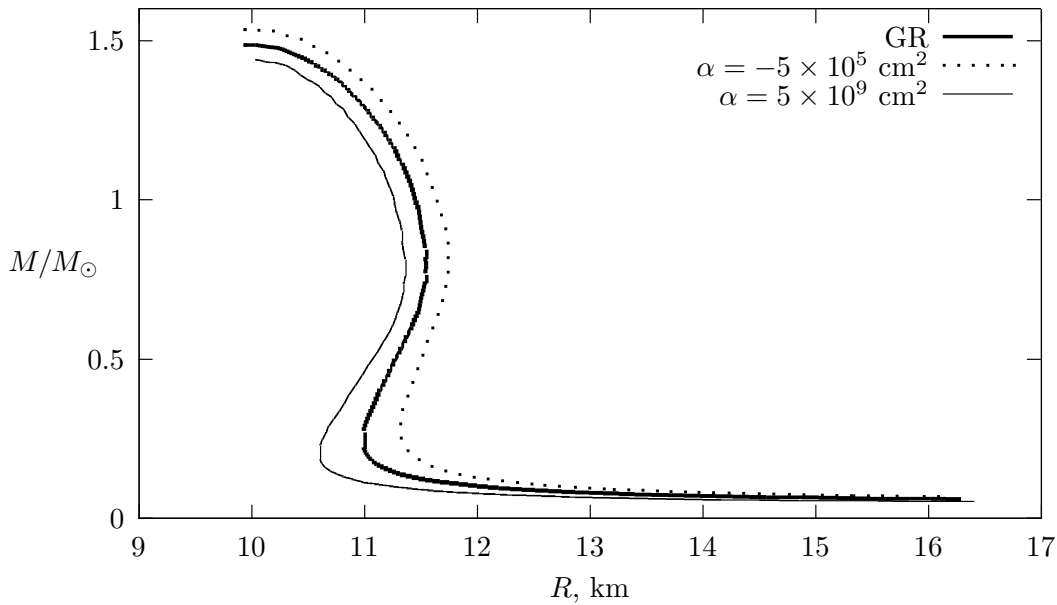}\\
  \caption{The mass-radius diagram in model $f(R)=R+\alpha R^2$ for two values of $\alpha$ without magnetic field in comparison with GR.}
\end{figure}

\begin{figure}
  \includegraphics[scale=1.1]{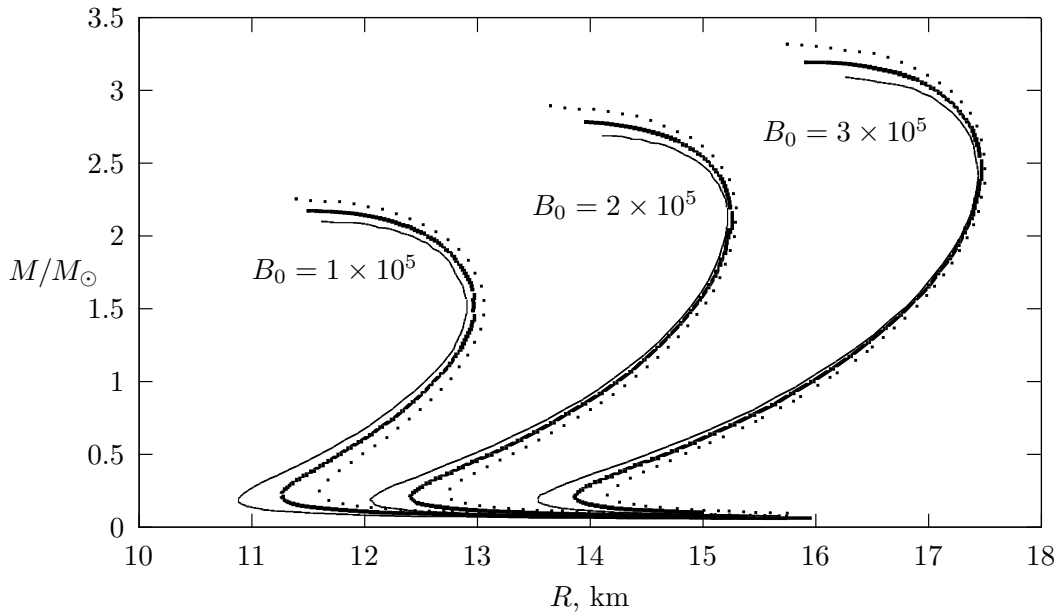}\\
  \caption{The mass-radius diagram in model $f(R)=R+\alpha R^2$ and in GR for slowly varying magnetic field ($B_{0}=1,\quad 2,\quad 3\times 10^5$).
  The cases $\alpha=-5\times 10^9$, $0$, $5\times 10^9$ cm$^2$
  correspond to dotted, thick and thin lines correspondingly.}
\end{figure}

\begin{figure}
  \includegraphics[scale=1.1]{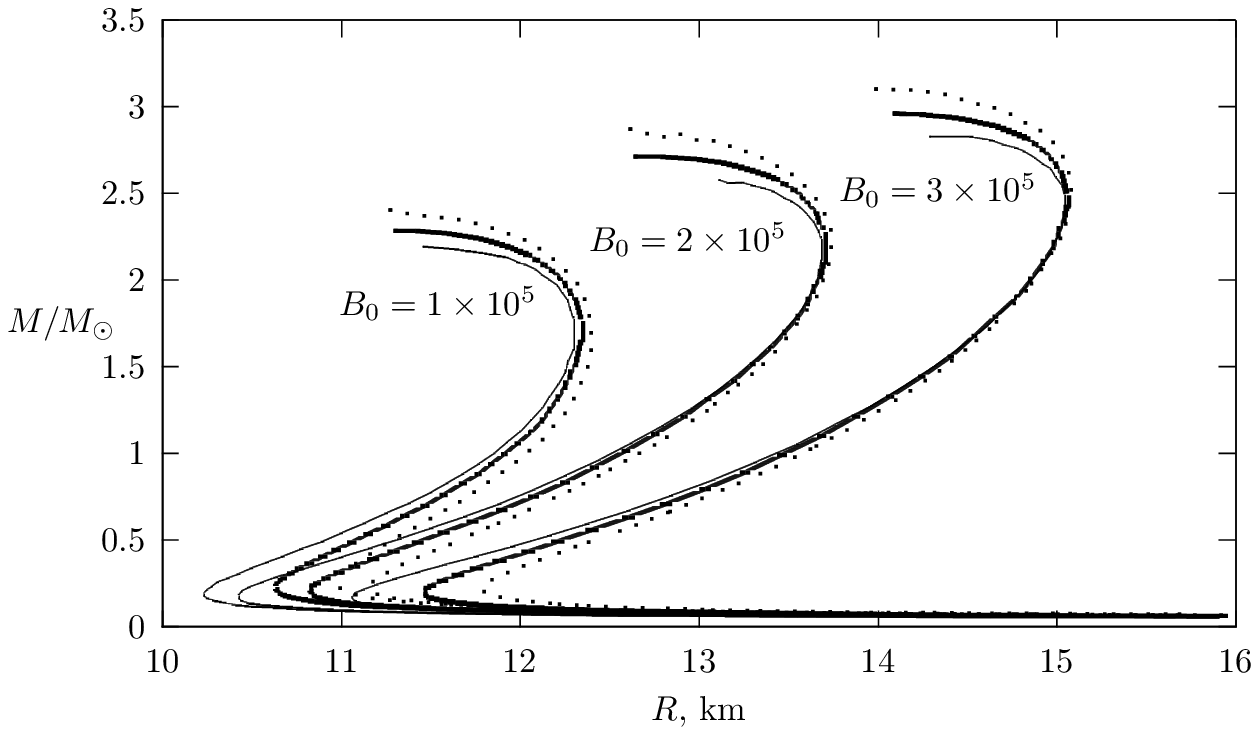}\\
  \includegraphics[scale=1.1]{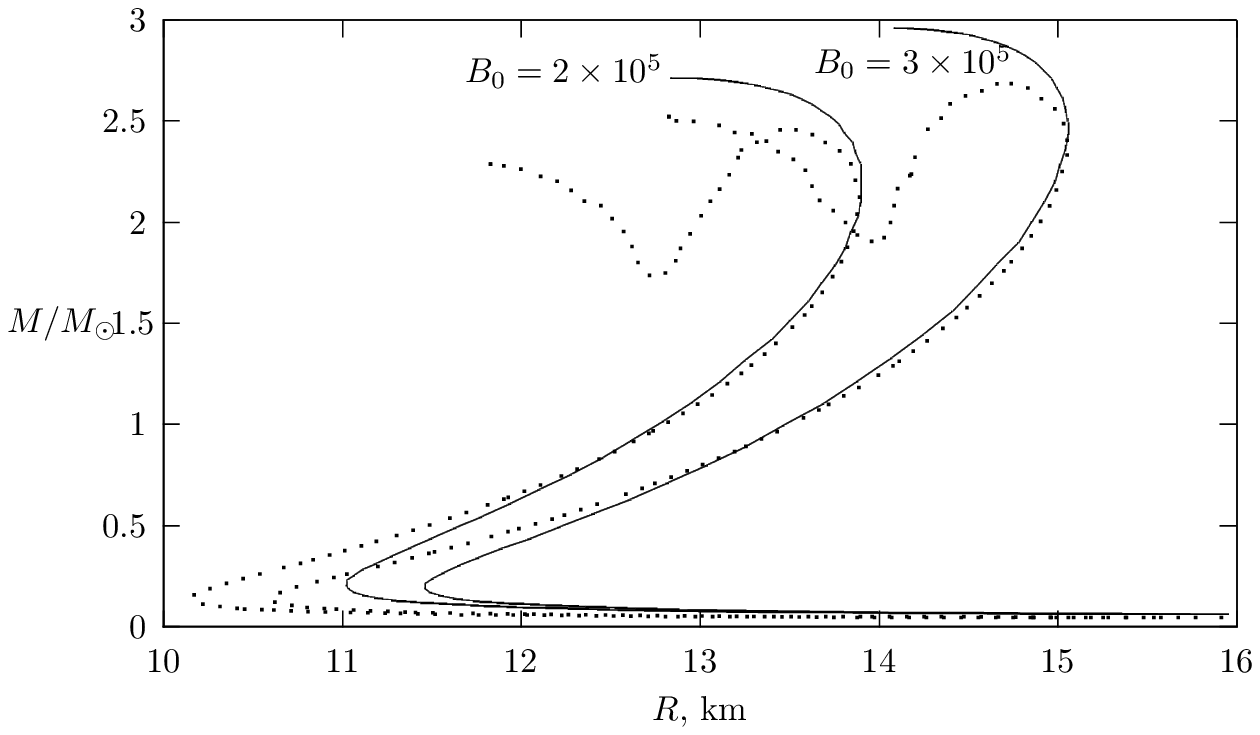}\\
  \caption{The Mass-Radius diagram in model $f(R)=R+\alpha R^2$ and in GR for fast varying field.
  On upper panel the cases $\alpha=-5\times 10^9$, $0$, $5\times 10^9$ cm$^2$
  correspond to dotted, thick and thin lines correspondingly. On lower panel the mass radius relation for $\alpha=10^{10}$ cm$^2$ is given (dotted lines).
  The second ``branch of stability'' with more compact (in comparison with GR) neutron stars exists.}
\end{figure}
\end{center}

\begin{table}
\label{Table5}
\begin{centering}
\begin{tabular}{|c|c|c|c|c|c|}
  \hline
  $B_{0}$,  & $\beta$, & $M_{max}$, & $R$,  & $E_{c}$, & $B_{c}$, \\
  $10^5$ & $r_{g}^{4}$ & $M_{\odot}$ & km & GeV/fm$^{3}$ & $10^{18}$ G \\
  \hline
   & 0 & 1.51 & 10.00 & 1.61 & 0 \\
  0 & $-50$ & 2.11 & 9.87 & 1.27 & 0 \\
   & $-75$ & 2.45 & 10.02 & 1.22 & 0 \\
   \hline
   & 0 & 2.21 & 11.69 & 1.17 & 3.38 \\
  1 & $-50$ & 2.70  & 11.07 & 1.67 & 4.09 \\
   & $-75$ &  3.10 &  10.97 & 1.81 & 4.19 \\
   \hline
      & 0 & 2.80 & 13.99 & 0.79 & 3.50 \\
  2 & $-25$ & 3.07 & 13.52 & 0.93 & 3.97 \\
    & $-50$ & 3.29 & 13.78 & 0.83 & 3.62 \\
   \hline
\end{tabular}
\caption{Neutron star properties using TW model for cubic gravity
for several values of $\beta$ (in units of $r_{g}^4=4.73\times
10^{21}$ cm$^4$) for slowly varying magnetic field.}
\end{centering}
\end{table}

\begin{table}
\label{Table6}
\begin{centering}
\begin{tabular}{|c|c|c|c|c|c|}
  \hline
  $B_{0}$,  & $\beta$, & $M_{max}/M_{\odot}$ & $R$,  & $E_{c}$, & $B_{c}$, \\
  $10^5$ & $r_{g}^{4}$ &  & km & GeV/fm$^{3}$ & $10^{18}$ G \\
  \hline
    & 0     & 2.32 & 11.50 & 1.17 & 3.96 \\
    & $-25$ & 2.73 & 11.10 & 1.27 & 4.14 \\
  1 & $-50$ & 3.14 & 11.19 & 1.17 & 3.96 \\
    & $-75$ & 3.64 & 11.18 & 1.17 & 3.96 \\
   \hline
    & 0     & 2.73 & 12.44 & 0.93 & 4.20 \\
  2 & $-25$ & 3.24 & 12.87 & 0.86 & 3.93 \\
    & $-50$ & 3.71 & 13.09 & 0.76 & 3.54 \\
   \hline
\end{tabular}
\caption{Compact star properties using TW model for cubic gravity
for several values of $\beta$ for fast varying magnetic field.}
\end{centering}
\end{table}

\begin{figure}
  \includegraphics[scale=1.1]{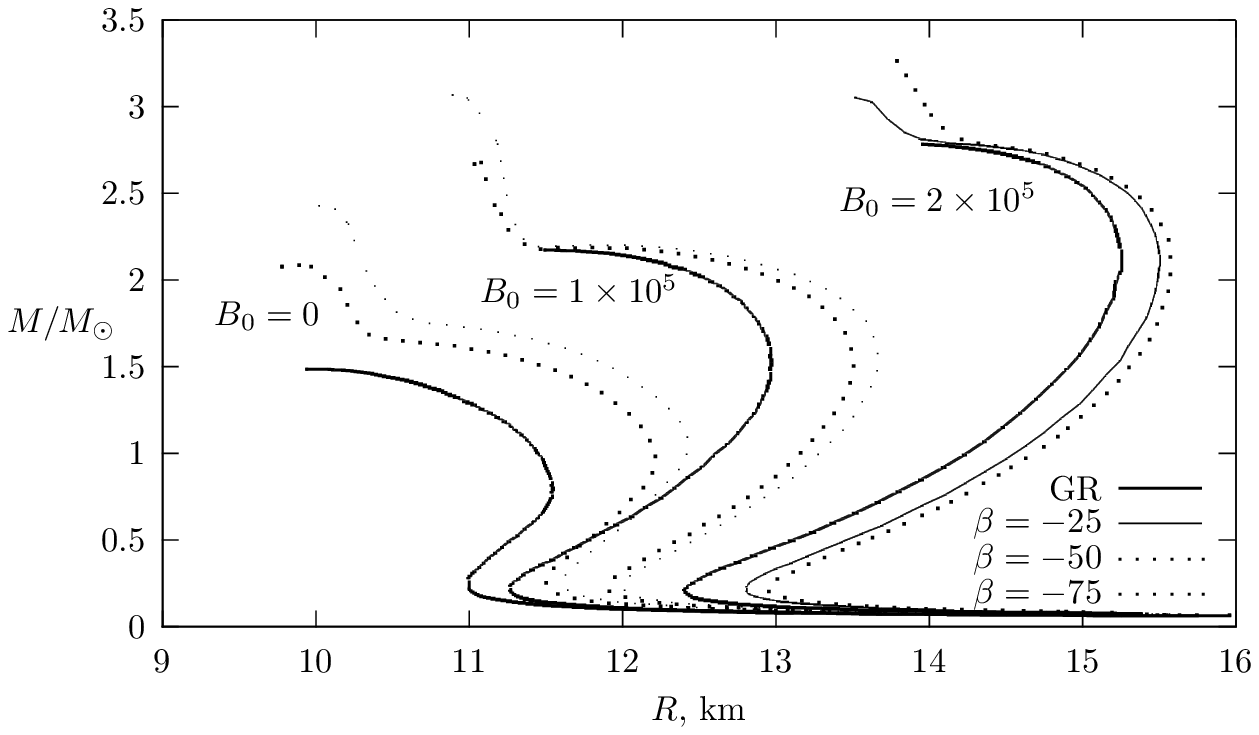}\\
  \includegraphics[scale=1.1]{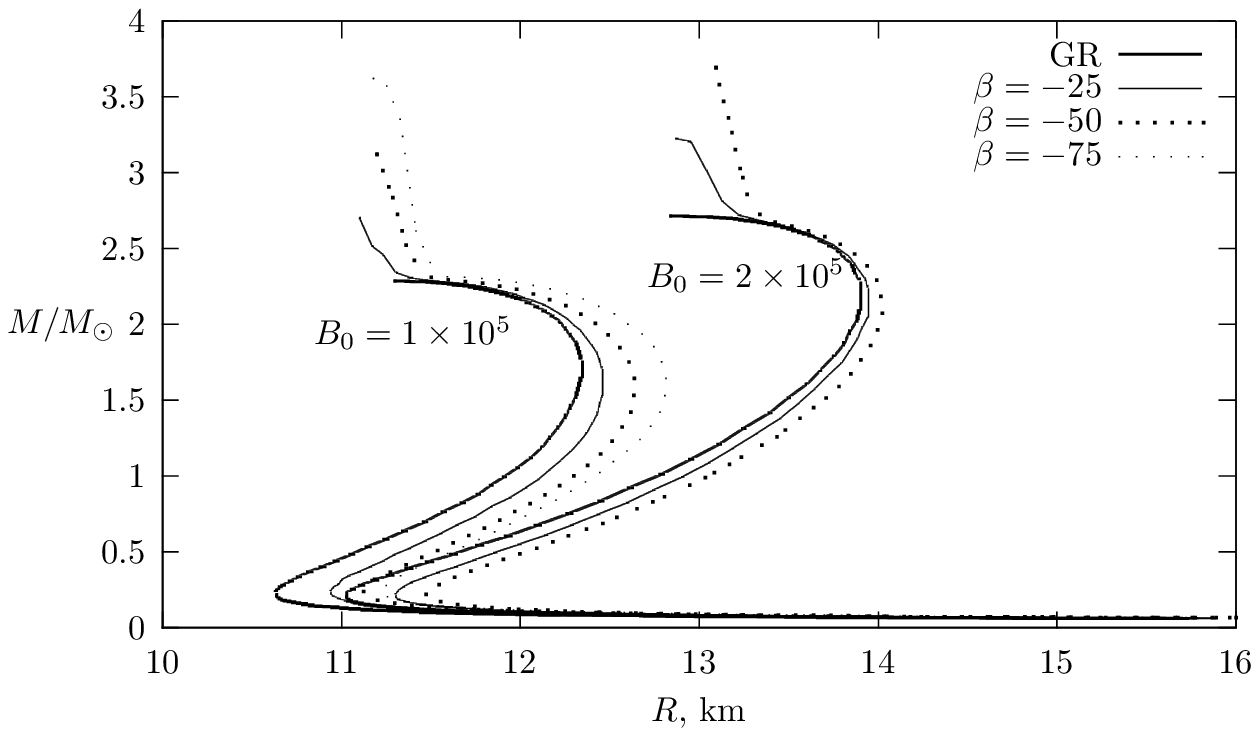}\\
  \caption{The mass-radius diagram in model $f(R)=R+\beta R^3$ and in GR for slowly (upper panel) and fast (lower panel) varying field. One can see that
  for the deviation of M-R relation from GR is smaller for larger values of
  $B_{0}$.}
\end{figure}

\section{The cases of quadratic and cubic curvature corrections}

Let us  firstly  take into account   models with quadratic gravity corrections, that is
\begin{equation}
f(R)=R+\alpha R^2.
\end{equation}
Neutron stars with strong magnetic field in quadratic gravity
was considered in \cite{EKSI} for relatively stiff EoS based on
model with five meson fields. We consider the quadratic gravity
case for EoS based on the above described model. For our calculations,  the
Typel-Wolter (TW) parametrization is used.

Let us  note the following feature: for $B=0$, the mass of neutron star
increases with decreasing $\alpha$  for the
various radii (see Fig. 1). For strong magnetic field,  the $M-R$
relation for $M>0.7M_{\odot}$ differs from such in GR only for masses  close to the maximal one (see Figs. 2,
3). Another interesting feature appears for fast varying field. At
high central densities,  a second ``branch'' of stability can
exists (Fig. 3, lower panel).

It is interesting to note that similar effects take place for
non-magnetic neutron stars in the framework  of a model like  $f(R)=R+\alpha
R^2(1+\gamma R)$ \cite{Astashenok}. The stabilization of star
configurations occurs thanks  to the cubic term.

The maximal masses and corresponding radii are given in Tables II,
III for some values of $\alpha$ and $B_{0}$. The maximal value of
central density (and therefore magnetic field) decreases with
increasing $\alpha$.

The parameters for compact (in comparison with GR) neutron stars
on second ``branch of stability'' are given in Table IV.

 For modified gravity with cubic term, $f(R)=R+\beta R^{3}$,  the
maximal value of neutron star mass for given EoS increases for
$\beta<0$ (Fig. 4). Some results are  given in Tables V, VI. The
maximal mass of neutron star can exceed $3M_{\odot}$. One can note
that  stars with magnetic field  and  cubic curvature corrections  result stable for  central energy density close to $\sim 1.8$
GeV/fm$^3$.

In principle, calculations show that, for EoS based on GM2-GM3 parameterizations,
we have   similar results for models with
$f(R)=R+\alpha R^2$ and $f(R)=R+\beta R^3$. For more stiff EoS the
deviation from GR is larger.

\section{Conclusions and perspectives}

We   presented   neutron star models with strong
magnetic fields in the framework of  power-law $f(R)$ gravity models. For
describing  dense matter in magnetic field, a model with
baryon octet interacting through $\sigma$$\rho$$\omega$-fields is
used.

Although the softening of nucleon EoS, due to hyperonization,
leads to the decrease of the upper limit mass of neutron star, the
strong magnetic field can increase considerably the maximal mass
of star.

In particular, we investigated the effect of strong magnetic field in  models of
quadratic, $f(R)=R+\alpha R^2$, and cubic, $f(R)=R+\alpha R^3$,
gravity. For large fields, the $M-R$ relation differs considerably
from such in GR only for stars with masses close
to maximal. Another interesting feature is the possible existence of
more compact stable stars with extremely large fields ($\sim
6\times 10^{18}$ G instead $\sim 4\times 10^{18}$ G in GR) in central regions of star. Due to the cubic term, the
significant increasing of maximal mass ($M_{max}>3M_{\odot}$) is
possible. The central energy density can exceed $\sim 1.8$
GeV/fm$^3$.

However, it is worth stressing  that  the $f(R)$ models considered here  can be  related to the presence of  strong gravitational fields where  higher order curvature terms can emerge.  Their  origin is related to the effective actions of quantum field theory formulated in curved spacetime \cite{buchbinder, birrel}.  In the extreme field  of neutron stars, it is realistic   supposing the emergence of  curvature corrections that  improve the pressure effects and could explain   supermassive self-gravitating systems.

As a next step, we will consider  models of self-bounded quark
stars and hybrid stars with quark cores. The EoS for quark matter
(without magnetic field) is close to $p\sim \frac{1}{3} \rho c^2$
and therefore, in the framework of perturbative approach, the deviations
from GR occur only for very large values of
$\alpha$ in comparison with the above considered in quadratic
gravity. However,  for large magnetic fields,  considerable effects can
be induced on EoS and therefore the modified gravity effects can
appear.

\acknowledgments

This work is supported in part by projects 14-02-31100 (RFBR,
Russia) (AVA), by MINECO (Spain), FIS2010-15640 and by MES project
TSPU-139 (Russia) (SDO). SC is supported by INFN ({\it iniziative
specifiche} TEONGRAV and QGSKY).

\end{document}